# Four-dimensional direct detection with Jones space optical full-field recovery

*Qi Wu, Yixiao Zhu\*, Hexun Jiang, Mengfan Fu, Yikun Zhang, Qunbi Zhuge, and Weisheng Hu*

Q. Wu, Y. Zhu, H. Jiang, M. Fu, Y. Zhang, Q. Zhuge, W. Hu

State Key Lab of Advanced Communication Systems and Networks, Department of Electronic Engineering, Shanghai Jiao Tong University, Shanghai, 200240, China
E-mail: yixiaozhu@sjtu.edu.cn

Y. Zhu, Q. Zhuge, W. Hu

Peng Cheng Laboratory, Shenzhen 518055, China

## Abstract

Data centers, the engines of the global Internet, are supported by massive high-speed optical interconnects. In optical fiber communication, the classic direct detection obtains only the intensity of the optical field, while the coherent detection counterpart utilizes both phase and polarization diversities at the expense of beating with a narrow-linewidth and high-stable local oscillator (LO). Herein, we propose and demonstrate a four-dimensional Jones space optical field recovery (4-D JSFR) scheme without LO. The information encoded on the intensity and phase of both polarizations can be captured by the polarization-diversity full-field receiver structure and subsequently extracted through deep neural network-aided field recovery. It achieves similar electrical spectral efficiency as standard intradyne coherent detection. The fully recovered optical field can extend the transmission distance beyond the power fading limitation induced by fiber chromatic dispersion. Furthermore, the LO-free advantage makes 4-D JSFR suitable for monolithic photonic integration, offering a spectrally efficient and cost-effective candidate for large-scale data center applications. Our results could motivate a fundamental paradigm shift in the optical field recovery theory and future optical transceiver design.

# 1. Introduction

Lightwave plays a critical role in data centers for information transmission and exchange with large capacity and extended reach [1-3]. Optical communication systems can be categorized as direct detection and coherent detection, depending on the existence of the local oscillator (LO) [4]. Since the birth of laser source, intensity modulation with direct detection (IM-DD) has been widely implemented as a low-cost solution [5-6]. However, the square-law detection of the photodiode (PD) limits not only the encoded degree of freedom to one dimension but also the transmission distance due to the power fading effect induced by fiber chromatic dispersion [7]. On the other hand, with the aid of LO beating, both phase and polarization diversities are exploited in coherent detection, bringing quadrupled electrical spectral efficiency and capacity [8-10]. After introducing digital signal processing (DSP) techniques, the linear optical-to-electrical mapping allows fiber impairment compensation in the coherent system and thus endows it with transoceanic and terrestrial transmission capability [11-15]. Nevertheless, narrow-linewidth laser sources are desired to suppress the phase noise [16], which prevents its massive deployment in cost-sensitive and fast-evolving short-reach scenarios represented by data-center optical interconnects.

To fully reach the high spectral efficiency advantage of coherent detection while maintaining the LO-free property of direct detection, novel receiver structures are highly desirable with phase diversity and polarization diversities, and field recovery ability. Pioneer efforts have been reported, known as the differential self-coherent receiver, which relies on beating signal with its delay [17]. Recently, a variety of carrier-assisted self-coherent schemes are gaining growing attention, including the Kramers-Kronig receiver [18], carrier-assisted differential detection [19], and asymmetric self-coherent detection [20]. They can reconstruct the optical field from the linear signal-carrier beating component. As such, phase diversity is successfully adopted without LO.

In addition to phase retrieval, the polarization fading (PF) phenomenon is a long-standing fundamental obstacle to accessing polarization diversity [21]. Unlike coherent detection beating with controllable LO at the receiver, direct detection schemes deliver an optical carrier together with the signal at the transmitter. The random polarization rotation during fiber transmission would lead to the absence of carrier on one polarization as the worst case, which loses the linear signal replica as a singularity. Such singularity cannot be compensated with multi-input-multi-output (MIMO) equalization in direct detection systems. Over the last decade, the Stokes vector receiver has been extensively studied as a promising solution to combat polarization fading [21-28]. However, the common phase of the two orthogonal polarizations is lost in the real-valued three-dimensional Stokes space. Attempts are made to further exploit the fourth dimension as inter-polarization differential phase using an additional optical hybrid with carrier-less signals [25, 29]. Alternatively, the modified Gerchberg-Saxton algorithm [30, 31] is recently introduced in optical communications. The dual-polarization optical field is reconstructed by digital propagation between dispersive projections but slowly arrives at convergence after hundreds of iterations [32].

Here we propose and demonstrate a four-dimensional Jones-space optical field recovery (4-D JSFR) scheme with the advance in both receiver structure and deep neural network (DNN)-aided field recovery. The 4-D JSFR receiver consists of a 3×3 optical coupler to ensure the satisfied carrier-to-signal power ratio (CSPR) threshold of both polarizations and three two-branch functional units that are responsible for both in-phase and quadrature information retaining. Furthermore, we adopt a DNN to simultaneously realize optical field recovery, polarization demultiplexing, and channel-matched impairment compensation. As a proof-of-concept experiment, we demonstrate the transmission of 206.2 Gb/s net rate with polarization-division-multiplexed probabilistically-shaped 64-ary quadrature amplitude modulation (PS-64QAM) over 80-km single-mode fiber (SMF). It has achieved the highest net electrical spectral efficiency of 11.36 b/s/Hz among the four-dimensional direct detection systems to our best knowledge. The proposed 4-D JSFR can approach the spectral efficiency of the coherent system without using an extra LO and potentially provide quadrupled capacity over IM-DD system. The concept of JSFR makes homodyne-like self-coherent detection possible for high-speed and cost-sensitive data-center interconnects.

## 2. Results

### 2.1. Concept of Jones-space optical full-field recovery

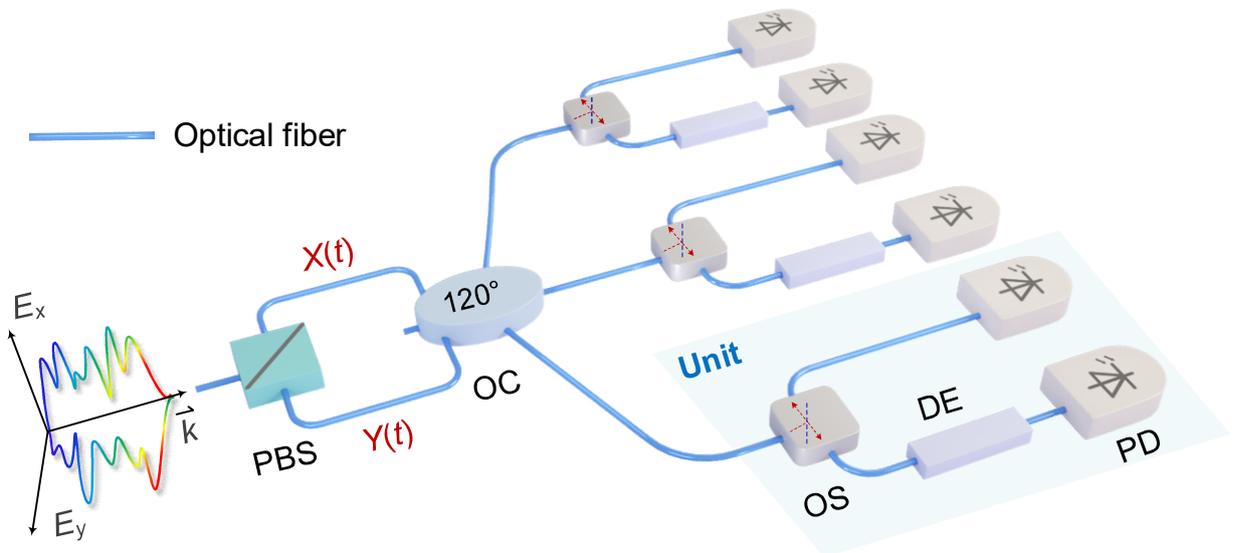

**Figure 1.** Polarization-diversity full-field receiver structure. PBS: polarization beam splitter; OC: 3×3 optical coupler; OS: optical splitter; DE: dispersive element; PD: photodiode. $E_x$ and $E_y$ are the orthogonal polarization directions.

Figure 1 shows the structure of the proposed polarization-diversity full-field receiver to detect the polarization-division-multiplexed (PDM) complex-valued double sideband signals. The minimum functional units in this receiver, consisting of two branches with dispersion diversity, are responsible for retaining the in-phase and quadrature information. The dispersion element (DE) acts like an all-pass filter having a time-domain impulse response of $h(t)$. After excluding the direct-current components, the detected photocurrents ($i_1(t)$ and $i_2(t)$) from two branches can be written as

$$i_1(t) - \langle i_1(t) \rangle = |C + S(t)|^2 - C^2 = 2C \cdot I(t) + |S(t)|^2 \tag{1}$$

$$i_2(t) - \langle i_2(t) \rangle = |\{C + S(t)\} \otimes h(t)|^2 - C^2 = 2C \cdot \left[ I(t) \otimes h_I(t) - Q(t) \otimes h_Q(t) \right] + |S(t) \otimes h(t)|^2 \tag{2}$$

where $C$ and $S(t)$ represent the optical field of the continuous-wave carrier and information-bearing complex-valued signal, respectively. '$\otimes$' stands for the convolution operation. The in-phase and quadrature components of $S(t)$ are denoted as $I(t)$ and $Q(t)$. The time-domain impulse response of DE can be decomposed into an in-phase part $h_I(t)$ and a quadrature part $h_Q(t)$ correspondingly. The 1st- and 2nd-terms on the right-hand side of both Eq. (1) and Eq. (2) correspond to the signal-carrier beating (namely the linear signal replica) and nonlinear signal-signal beat interference (SSBI), respectively. Specifically, Eq. (1) reveals the fact that direct detection can capture only the in-phase component $I(t)$ with respect to the optical carrier, while the quadrature information $Q(t)$ is completely lost. Thanks to the DE, the quadrature information is mixed with the in-phase part and captured in the dispersed branch in Eq. (2). Therefore, the linear combination of Eq. (1) and Eq. (2) can fully recover the complex-valued optical field once the SSBI impairment is compensated. In carrier-assisted self-coherent systems, CSPR has been considered a critical parameter for SSBI cancellation. The SSBI impairment can be regarded as a perturbation assuming the CSPR value exceeds the threshold $C_{th}$. Thus, it is possible to apply various digital linearization algorithms including iterative reconstruction [33, 34], Gerchberg-Saxton algorithm [35], and deep learning [36] to eliminate SSBI. In our work, a deep neural network is chosen due to its advantages in accurate channel parameter estimation and nonlinear fitting.

For PDM signals reception, without loss of generality, we assume the optical carrier is linearly polarized, located at $\pi/4$ angle between X- and Y-polarization. In principle, the scheme is still available for an arbitrary state of polarization (SOP) of the carrier case (see Supplementary Note 2 for detailed derivation). As such, the Jones vector of the transmitted carrier-assisted dual-polarization optical field is denoted as [$C+S_x(t)$, $C+S_y(t)$], where $S_x(t)$ and $S_y(t)$ are independent complex-valued signals with equal power on the X- and Y-polarizations, respectively. $C$ is the continuous wave. In such a dual-polarization case, the CSPR is defined as the power ratio between the two carriers ($P_C$) and the PDM signals of two polarizations ($P_{S_x}$, $P_{S_y}$), shown in Eq. (3).

$$CSPR = 10 \lg \frac{2 \cdot P_C}{P_{S_x} + P_{S_y}} = 10 \lg \frac{P_C}{P_{S_x}} \qquad (3)$$

Generally, the polarization effect in short-reach optical communications can be modeled in Jones space as a 2×2 frequency-independent unitary matrix [2]. After the transmission in SMF, the received Jones vector can be written as

$$\begin{bmatrix} X(t) \\ Y(t) \end{bmatrix} \triangleq \begin{bmatrix} \cos\alpha e^{j\theta} & -\sin\alpha \\ \sin\alpha & \cos\alpha e^{-j\theta} \end{bmatrix} \begin{bmatrix} C + S_x(t) \\ C + S_y(t) \end{bmatrix} \qquad (4)$$

where $\alpha$ and $\theta$ are the random polar angle and differential phase between X- and Y-polarization, respectively. Note that since the carriers on X- and Y-polarization depend on the unitary matrix, the CSPR condition of each polarization cannot be satisfied in the polarization fading case, which hinders the direct reconstruction of $X(t)$ and $Y(t)$ from the photocurrents.

In order to overcome the polarization fading problem, we harness a 3×3 optical coupler to mix $X(t)$ and $Y(t)$. Subsequently, three parallel two-branch functional units are used to detect the three optical field outputs of the optical coupler (see Supplementary Note 1). The critical point to ensure optical field recovery is that the CSPR values of both polarizations should always exceed the threshold, namely $C_{th}$, for all the SOP situations. Using a polarization-independent transformation matrix, we can digitally reconstruct the photocurrents of optical fields $X(t)+Y(t)$, $X(t)-Y(t)$, $X(t)+jY(t)$, and $X(t)-jY(t)$ (see Supplementary Note 1 for detailed theoretical derivation). The CSPR values of these four optical fields are calculated as

$$\begin{aligned} CSPR_{X(t)+Y(t)}(\alpha,\theta) &= \left(2\cos^2\alpha\cos^2\theta\right) \cdot C_{th} \\ CSPR_{X(t)-Y(t)}(\alpha,\theta) &= \left(2 - 2\cos^2\alpha\cos^2\theta\right) \cdot C_{th} \\ CSPR_{X(t)+jY(t)}(\alpha,\theta) &= \left(1 + 2\cos^2\alpha\cos\theta\sin\theta\right) \cdot C_{th} \\ CSPR_{X(t)-jY(t)}(\alpha,\theta) &= \left(1 - 2\cos^2\alpha\cos\theta\sin\theta\right) \cdot C_{th} \end{aligned} \qquad (5)$$

Equation (5) is mathematically beautiful. Although the four CSPR values vary with $\alpha$ and $\theta$, it can be found that $CSPR_{X(t)+Y(t)} + CSPR_{X(t)-Y(t)} = CSPR_{X(t)+jY(t)} + CSPR_{X(t)-jY(t)} = 2 \cdot C_{th}$. The energy conservation law of the carrier always holds within the pairs of {$CSPR_{X(t)+Y(t)}$, $CSPR_{X(t)-Y(t)}$} and {$CSPR_{X(t)+jY(t)}$, $CSPR_{X(t)-jY(t)}$}. It indicates that there are always two CSPR values larger than the threshold $C_{th}$ for the perturbation assumption, regardless of polarization rotation. After parallel optical field reconstruction on the linearly combined photocurrents, one in each pair can be reconstructed accurately. Polarization demultiplexing is further available thanks to the unitary property of the transformation matrix (see Eq.(S7) in Supplementary Note 1). Therefore, the polarization fading obstacle is finally eliminated with the polarization-diversity full-field receiver structure and the Jones space field recovery methodology. At this time, both phase and polarization diversities are accessible.

## 2.2. Experimental setup for four-dimensional direct detection

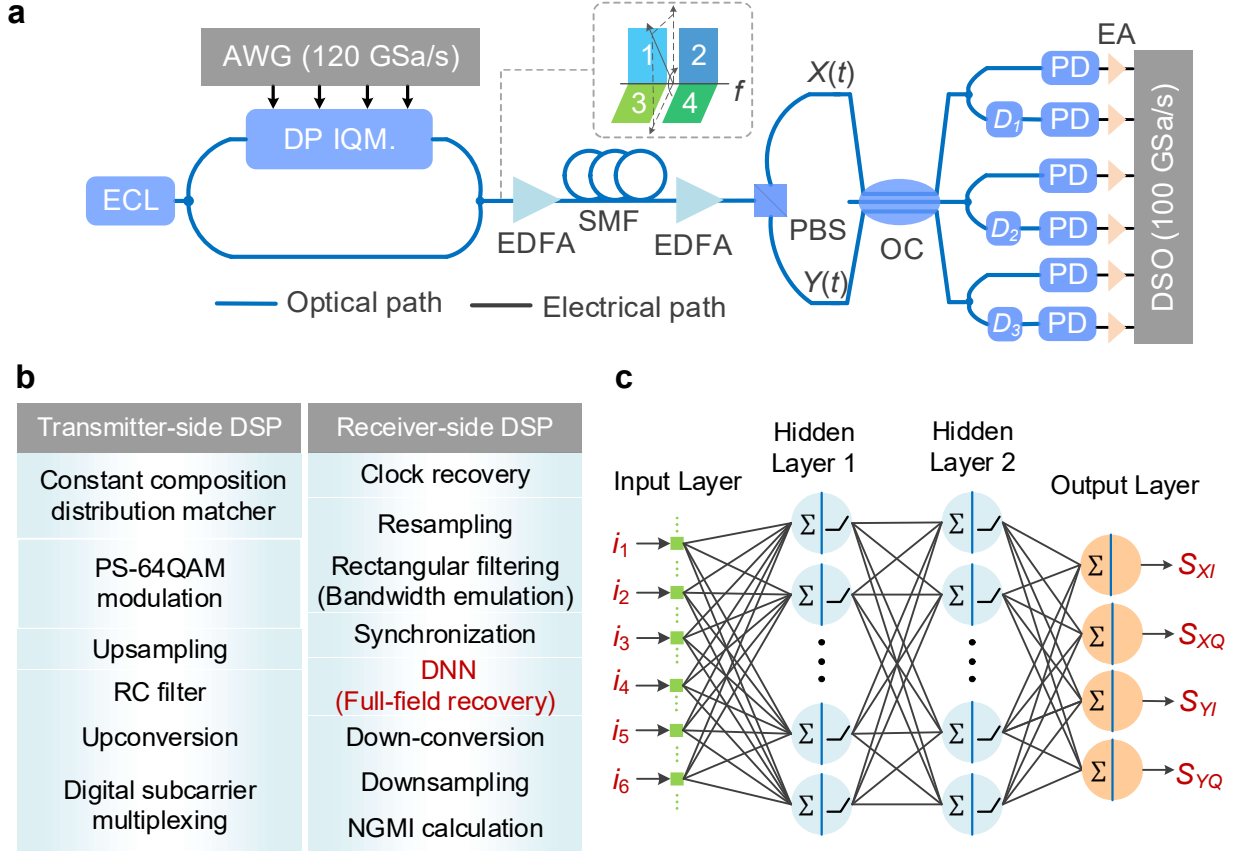

**Figure 2.** (a) Experimental setup. ECL: external cavity laser; AWG: arbitrary waveform generator; DP IQM.: dual-polarization IQ modulator; EDFA: erbium-doped fiber amplifier; SMF: single-mode fiber; PBS: polarization beam splitter; OC: 3×3 optical coupler; PD: photodetector; EA: electrical amplifier; DSO: digital storage oscilloscope. $D_1$, $D_2$ and $D_3$ are the three dispersive elements with a dispersion value of -300 ps/nm. $i_k$ ($k$=1~6), are the six received photocurrents. $S_{XI}$, $S_{XQ}$, $S_{YI}$, and $S_{YQ}$ are the in-phase and quadrature components of transmitted polarization-multiplexed signals. (b) DSP stacks. RC: raised cosine; MIMO: multi-input-multi-output; DNN: deep neural network; NGMI: normalized generalized mutual information. (c) Structure of deep neural network.

Figure 2(a) shows the experimental setup for the proposed 4-D JSFR scheme. The procedures of the offline DSP for signal modulation, optical field recovery, and signal demodulation are shown in Fig. 2(b). At the transmitter, an arbitrary waveform generator operating at 120 GSa/s is used to generate 15 GBd PS-64QAM symbols shaped by raised cosine filter with a roll-off factor of 0.01. We adopt digital subcarrier multiplexing to avoid the singularity of the receiver transfer function [36]. The four independent signals are up-converted to the intermediate frequency of 10.5 GHz to reserve a 3-GHz guard band. The electrical signals are loaded into an integrated dual-polarization IQ modulator to modulate the light from the external cavity laser (ECL) centered at 1551.06 nm. Another branch from ECL is employed as the optical carrier with uncontrolled SOP. The modulated optical field is amplified by an erbium-doped fiber amplifier (EDFA) and then launched into the fiber link. The launch power is optimized as 8 dBm. We index the four signal bands by their polarization and frequency, as shown in the inset of Fig. 2(a).

The optical signal is first boosted to 18 dBm after 80-km SMF transmission and then fed into the polarization-diversity full-field receiver for 4-D JSFR. Here, the 3×3 optical coupler and three two-branch functional units are used to eliminate the polarization fading effect and recover the single-polarization optical field, respectively. The applied dispersion values of three dispersive elements are set as about -300 ps/nm. After the photoelectric conversion, a 100 GSa/s real-time digital storage oscilloscope with six synchronized channels, is used to capture the electrical waveforms for offline DSP. At the receiver-side DSP, the electrical waveforms are resampled to 60 GSa/s and then digitally filtered to emulate a receiver with a rectangular bandwidth of 18.15 GHz. After the frame synchronization, the six digital waveforms are fed into the four-layer deep neural network (see Materials and methods) to perform optical full-field recovery, polarization demultiplexing, and channel equalization simultaneously. The reconstructed dual-polarization optical signal fields are demodulated to the baseband and then down-sampled. The normalized generalized mutual information (NGMI) of signal bands 1-4 are calculated using the bit-wise log-likelihood ratio [37]. Additionally, we harness the transfer learning method [38] to reduce the redundancy of preambles in each frame, and thus only 3000 symbols are required for the convergence of DNN (see Supplementary Note 4). The rest of the 57000 symbols are used as payload to calculate NGMI. A detailed description and illustration of the experimental setup are available in Supplementary Note 3.

## 2.3. Source entropy and CSPR optimization

In this experiment, we adopt probabilistic shaping [39] to approach the linear Shannon limit. Four independent PS-64QAM symbol sequences with a length of 60000 are generated based on the constant composition distribution matcher [40]. The source entropy is optimized as 4.4 bits/symbol to maximize the achievable information rate in terms of generalized mutual information (GMI) for this system. The transmitted probability distribution of the PS-64QAM constellation with a shaping parameter $\beta$=1.1976 is shown in Fig. 3(a). The probabilities of the outer constellation points are relatively reduced to enlarge the Euclidean distance under average power constraint.

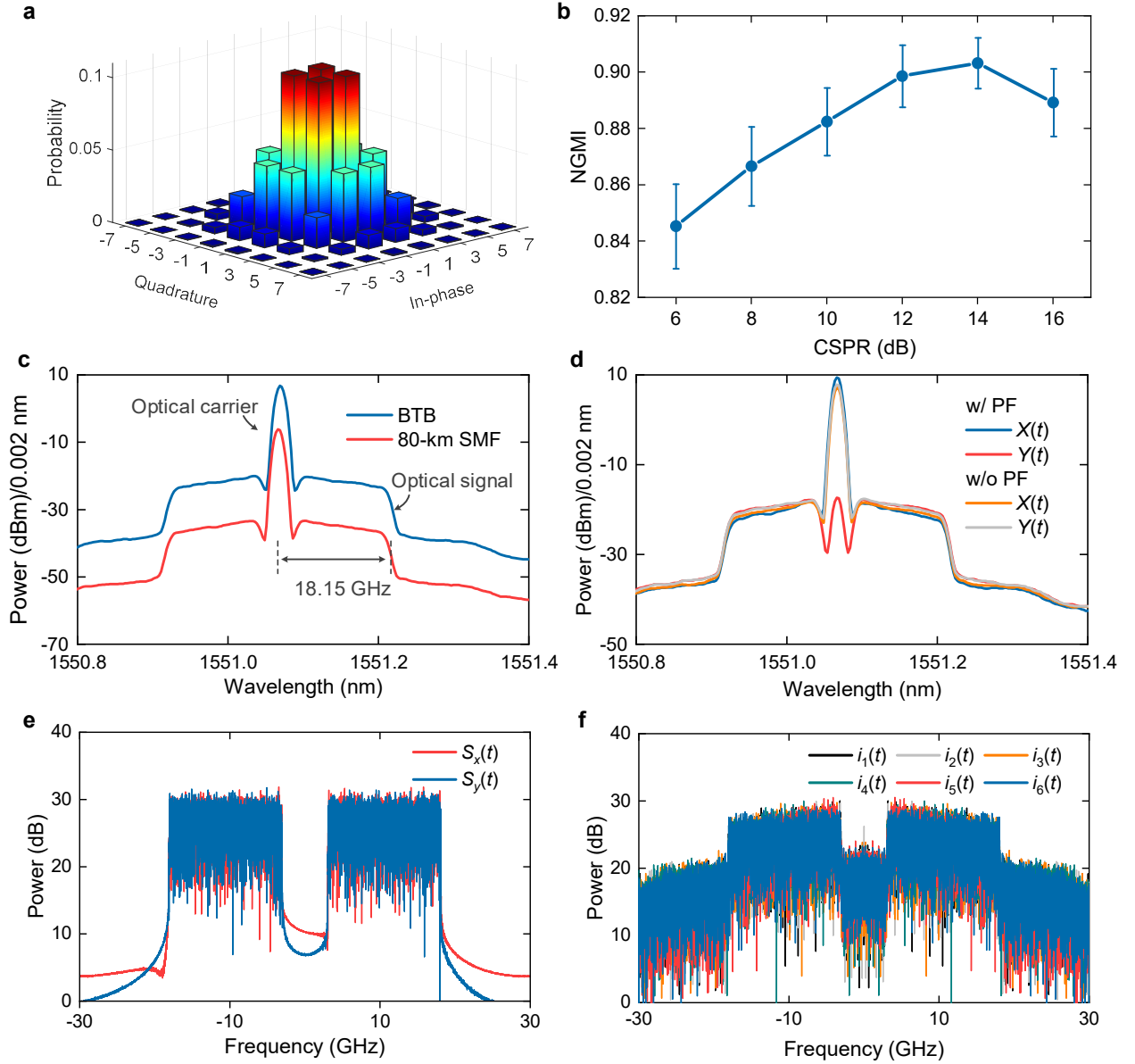

**Figure 3.** (a) Symbol probability distribution for the PS-64QAM. (b) Measured NGMI versus CSPR in the 80-km transmission case. (c) Optical spectra of PDM optical signals in the BTB and 80-km SMF case. (d) Optical spectra of X- and Y-polarization signals with and without polarization fading phenomenon. PF: polarization fading. (e) Electrical spectra of the transmitted PDM signals. (f) Electrical spectra of the six received photocurrents. $i_k(t)$ ($k$=1~6), are the six received photocurrents.

Additionally, CSPR is a critical factor needed to be optimized for the JSFR scheme in the presence of nonlinear SSBI impairment. A higher CSPR degrades the effective optical signal-to-noise ratio (OSNR), while a lower CSPR cannot satisfy the perturbation assumption for accurate nonlinear SSBI compensation. We thus sweep the CSPR from 6 to 16 dB to identify the optimal value for 80 km transmission, shown in Fig. 3(b). It shows that the optimal CSPR is about 14 dB for dual-polarization optical field recovery. Under this CSPR, the received optical spectra in the back-to-back (BTB) and 80-km SMF transmission case at a resolution of 0.002 nm are shown in Fig. 3(c). It can be found that half of the optical signal bandwidth is ~0.145 nm. It is equivalent to ~18 GHz bandwidth around 1550 nm wavelength, containing 15.15 GHz signal bandwidth and a 3 GHz guard band. Figure 3(d) plots the comparisons of optical spectra of both polarizations with and without polarization fading phenomenon. It shows that the optical carrier power difference between X- and Y-polarizations is negligible without polarization fading. However, the optical field on one of two polarizations is carrier-less when polarization fading occurs, which will be complemented after mixing in the 3×3 optical coupler. Otherwise, the carrier-less polarization cannot be recovered using the principle described in Eq. (1) and (2) since the optical carrier is not strong enough for the perturbation assumption. The transmitted and received electrical spectra are shown in Fig. 3(d)-(e), respectively. The signal-carrier beating term contributes to the main part of the received signal spectra, and the nonlinear SSBI spans a much wider spectrum than the signal bandwidth. Fortunately, the required bandwidth of the JSFR receiver only needs to be larger than the occupied bandwidth of linear signal replica, as the settings in the receiver-side DSP.

## 2.4. Polarization rotation tolerance evaluation

We first test the system performance in two extreme cases, namely with and without polarization fading, to characterize its polarization rotation resilience. The transmission performance is indicated by NGMI, which can accurately predict post-forward error correction (FEC) performance independent of the source entropy of the PS-QAM signal [37]. Figure 4(a) investigates the NGMI as a function of OSNR in the BTB case. Using the optimal CSPR of 14 dB, the highest achievable OSNR is about 38.5 dB. By comparing the OSNR sensitivity curves, the NGMI performance with PF is slightly better than that without PF because the power of optical signal-carrier beating terms is maximized in the PF case, contributing to a higher electrical signal-to-noise ratio. In this experiment, we adopt the FEC with a code rate of 0.8402 (i.e., 19.02% overhead) and a corresponding NGMI threshold of 0.8798 [41] to remove the potential error floor. The OSNR penalty is 0.3 dB at this NGMI threshold, implying that the proposed 4-D JSFR has a good tolerance for polarization rotation. Note that the case of no activation function is plotted to represent the performance using the polarization-diversity full-field receiver and linear 6×4 MIMO equalization. Thus, Figure 4(a) also shows that the nonlinear activation function helps to eliminate the SSBI effectively generated in photoelectric conversion. In the low-OSNR regime, the nonlinear activation function has little impact on NGMI performance since the dominant impairment is optical noise instead of SSBI. Figure 4(b) depicts the OSNR sensitivity of the JSFR receiver after 80-km SMF, covering the transmission distance of the inter-data centers interconnection scenario. With an OSNR beyond 36 dB, the NGMI threshold is reached under different SOPs, indicating an error-free transmission. Compared with the BTB case, the field recovery ability of 4-D JSFR successfully brings the 80-km transmission penalty down to less than 0.3 dB. The corresponding GMI and bit error rate (BER) at an OSNR of 37 dB are labeled in this figure.

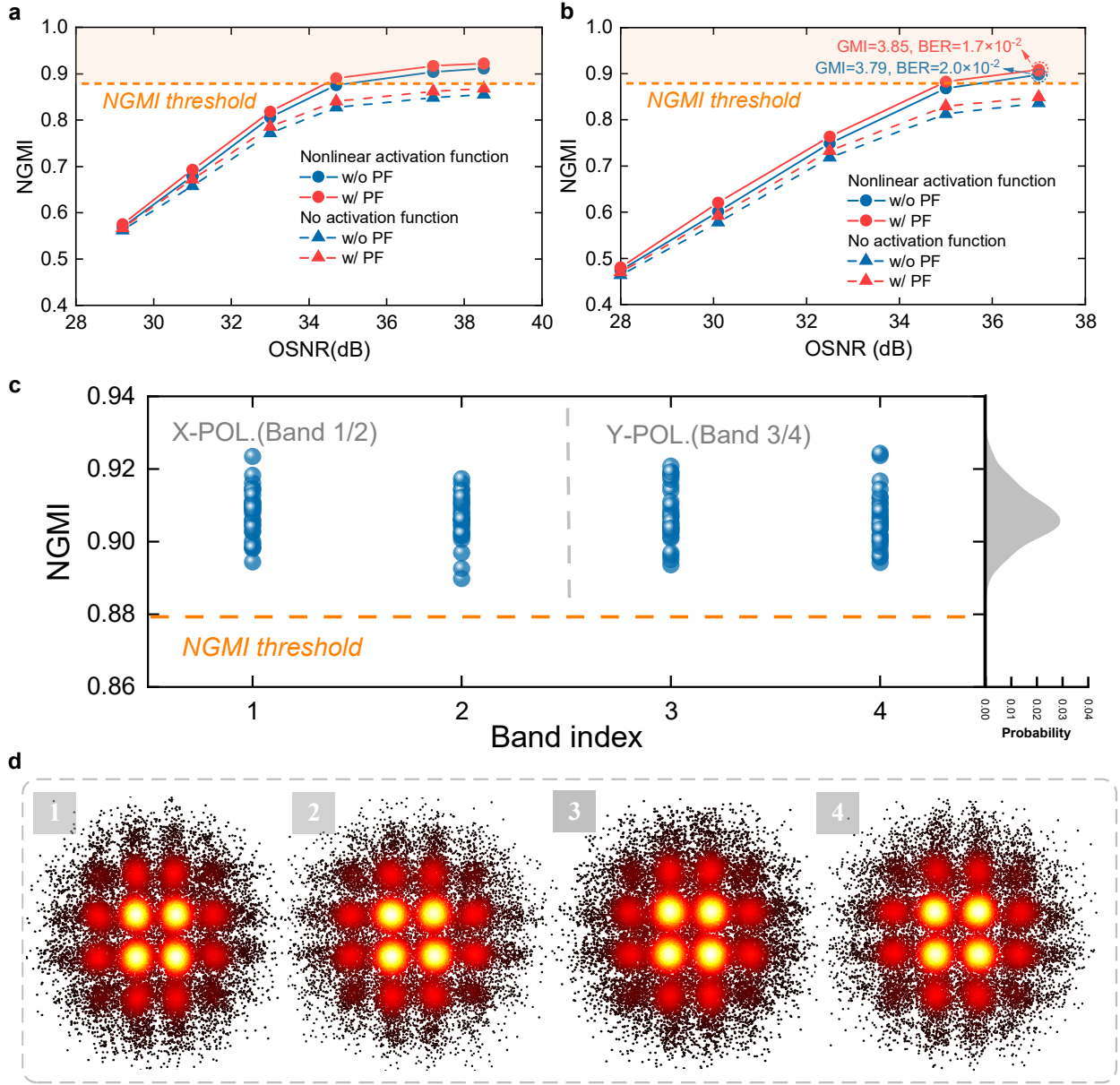

**Figure 4.** (a) Measured NGMI versus OSNR in the BTB case with and without nonlinear activation function in the BTB case. (b) Measured NGMI versus OSNR with and without nonlinear activation function in DNN after 80-km SMF transmission. GMI: generalized mutual information; BER: bit error rate. (c) NGMIs of the four signal bands over 30 measurements with randomly-varying polarization state. (d) Constellation of the four signal bands.

The practical system is expected to work consistently under any polarization situation. Here we measure the system performance under a randomly-varying polarization state. The results at 37 dB OSNR are displayed in Fig. 4(c). The NGMI distribution over 30 measurements is also presented as an inset on the right. Stable transmission performance can be verified from the negligible NGMI fluctuation within 0.03. The recovered constellations of four signal bands after 80-km SMF transmission are displayed in Fig. 4(d).

Here we highlight the achieved electrical spectral efficiency with 4-D JSFR in the experiment. The (net) data rate is calculated as (206.2) 263.8 Gbit/s. The rectangular electrical bandwidth of the 4-D JSFR receiver is 18.15 GHz. Thus, the net electrical spectral efficiency is 11.36(=206.2/18.15) b/s/Hz (see Supplementary Note 5). To the best of our knowledge, it achieves the highest electrical spectral efficiency among the existing four-dimensional direct detection systems.

## 3. Discussion

**Table 1.** Comparison of four-dimensional direct detection schemes

| Reference | Modulation format | Distance (km) | Bit rate (Gbit/s) | Required bandwidth of receiver (GHz) | Net electrical spectral efficiency (b/s/Hz) |
|-----------|-------------------|---------------|-------------------|--------------------------------------|---------------------------------------------|
| [32]      | QPSK              | 520           | 120               | 30.3                                 | 2.75                                        |
| [29]      | QPSK              | 1             | 119               | 36                                   | 3.09                                        |
| [25]      | 2-ring/8PSK       | 10            | 320               | 31.5                                 | 9.64                                        |
| This work | PS-64QAM          | 80            | 263.8             | 18.15                                | 11.36                                       |

The state-of-the-art four-dimensional direct detection schemes are summarized in Table 1. Compared with the schemes of [25] and [29], 80-km transmission distance can be supported with 4-D JSFR to fulfill the requirement of inter-connection between large-scale data centers, which reveals its capability of optical field recovery for fiber chromatic dispersion compensation. In addition, unlike the modified Gerchberg-Saxton algorithm that reconstructs the optical field from the quadratic signal-signal beating term of carrier-less signal [32], 4-D JSFR successfully extracts the optical field from the linear signal-carrier beating term. Therefore, the requirement bandwidth of receiver is almost halved with respect to signal bandwidth, making the electrical spectral efficiency comparable to intradyne coherent detection.

In our experiment, a Fiber Bragg grating-based dispersion compensation module is employed as the dispersive element. The dispersive elements can also be implemented using on-chip micro-ring resonators, optical all-pass filters, or Silicon Bragg gratings [42]. The critical consideration is to permit sufficient optical modulation bandwidth for ultra-wideband large-capacity transmission.

It is worth noting that the LO-less advantage makes the polarization-diversity full-field receiver structure especially attractive for the complementary metal oxide semiconductor (CMOS)-compatible silicon photonics platform [43-45]. Monolithic integration is within reach, especially for group IV element-based indirect band gap semiconductor material with low quantum efficiency.

The 4-D JSFR has addressed the polarization fading issue, which can be understood as polarization mode coupling in single-mode fiber. For spatial division multiplexing, the methodology can also be applied to eliminate the spatial mode/core coupling-induced fading phenomenon in the direct detection system. Thus, the JSFR scheme has the potential to exploit higher physical dimensions in few-mode [46], orbital-angular-momentum ring-core [47], or multi-core fiber [48].

This paper has described the concept and experimental results of a four-dimensional Jones-space optical full-field recovery transceiver to fully exploit the physical dimensions of light in single-mode fiber, including intensities and phases of both polarizations. The concept is verified by transmitting a 263.8-Gbit/s dual-polarization complex-valued double sideband signal over 80-km single-mode-fiber, achieving the highest electrical spectral efficiency of 11.36 b/s/Hz in the four-dimensional direct detection system. 4-D JSFR not only almost quadruples the conventional IM-DD system's electrical spectral efficiency but also enables digital compensation for chromatic dispersion and random polarization rotation. For the first time, the optical full-field is recovered in Jones space via direct detection. It provides a powerful and practical solution suitable for data center optical networks. The theoretical derivation and analysis to tackle polarization mode coupling issues can be extended to few-mode, multi-core fiber, and other communication channels with spatial mode division multiplexing. In addition to optical communications, the concept of phase retrieval and Jones-space optical full-field recovery schemes can also be applied to optical metrology, imaging, or sensing systems.

## 4. Materials and methods

### Training of DNN-based field linearization and SSBI elimination

The structure of the adopted transfer learning-assisted DNN is shown in Fig. 2(c). The gist of transfer learning is to reduce the required training epochs and symbols and hence realize fast remodeling (see Supplementary Note 4). For the 60000 PS-64QAM symbols generated by the constant composition distribution matcher, 3000 symbols (12000 samples) are considered the training dataset, and the rest of the 57000 symbols are the payload data used to calculate the pre-FEC NGMI. After being down-sampled to 60 GSa/s, the received six photocurrents are fed into the DNN for optical field linearization and SSBI elimination. The input vector of DNN is the received signal samples $i_k(n-L)$-$i_k(n+L)$, $k$=1~6, where $L$ is 35, and the output is the estimated sample sequences. The nodes of the input layer are 426(=71×6). Two hidden layers consisting of 512 and 256 nodes, respectively, and four output nodes of the output layer are the real and imaginary parts of the polarization-division-multiplexed training sequences. The rectified linear units (ReLU) function is adopted as the activation function thanks to its rapid converging speed. The Adam optimizer is employed to minimize the root-mean-squared error loss function of the DNN. In the training process, the Adam optimizer will optimize the weight and bias of DNN to match the fiber link-related parameters, especially the applied dispersion value of dispersive elements, which helps recover the single-polarization optical field. The initial learning rate is set as $1×10^{-3}$ for fast convergence and then decayed to $1.6×10^{-6}$ after 20 epochs for accurate optimization. After the training stage, the optimized network is switched to test the system performance.

## Acknowledgments.

This work was supported by National Natural Science Foundation of China (62001287, 62271305) and National Key R&D Program of China (2019YFB1803803).

## Author contributions.

Q.W. conceived the idea. Q.W., Y.Z.[1], H.J., M.F. and Y.Z.[2] contributed to the theoretical analyses and simulations. Q.W., and Y.Z.[1] performed the experiment and analyzed the data. Q.W. and Y.Z.[1] wrote the manuscript. All co-authors contributed by providing their valuable feedback and comments. Y.Z.[1] and W.H. supervised the project.

## Data availability.

Data underlying the results presented in this paper are not publicly available at this time but may be obtained from the authors upon reasonable request.

## Conflict of interest.

The authors declare that they have no conflict of interest.

Th3B.6.

# Supplementary Information for "Four-dimensional direct detection with Jones space optical full-field recovery"

*Qi Wu, Yixiao Zhu\*, Hexun Jiang, Mengfan Fu, Yikun Zhang, Qunbi Zhuge, and Weisheng Hu*

Q. Wu, Y. Zhu, H. Jiang, M. Fu, Y. Zhang, Q. Zhuge, W. Hu

State Key Lab of Advanced Communication Systems and Networks, Department of Electronic Engineering, Shanghai Jiao Tong University, Shanghai, 200240, China
E-mail: yixiaozhu@sjtu.edu.cn

Y. Zhu, Q. Zhuge, W. Hu

Peng Cheng Laboratory, Shenzhen 518055, China

Dated: December 30, 2022

# Contents

**Supplementary Note 1: Theoretical derivation of optical interference in the 3×3 optical coupler and transformation of photocurrents.**

After the dual-polarization optical field is split by the polarization beam splitter (PBS), two single-polarization optical fields, $X(t)$ and $Y(t)$, are acquired. In order to overcome the polarization fading phenomenon, we harness a 3×3 optical coupler (120° optical hybrid) to mix $X(t)$ and $Y(t)$. The three optical field outputs can be expressed as

$$\begin{bmatrix} a\,b\,b \\ b\,a\,b \\ b\,b\,a \end{bmatrix}\begin{bmatrix} X(t) \\ 0 \\ Y(t) \end{bmatrix} = \begin{bmatrix} aX(t)+bY(t) \\ bX(t)+bY(t) \\ bX(t)+aY(t) \end{bmatrix}, \tag{S1}$$

where $a$ and $b$ are $(2e^{j2\pi/9}+e^{-j4\pi/9})/3$ and $(e^{-j4\pi/9}-e^{j2\pi/9})/3$, respectively [1]. The three outputs are the optical fields of complex-valued double sideband signal. Then, we use three two-branch functional units behind the three outputs of the 3×3 optical coupler to detect the optical fields parallelly. The two-branch functional unit consists of one 1×2 optical splitter, one dispersive element ($D_i$) used for I/Q mixing, and two single-ended photodiodes. The received six photocurrents are denoted as

$$\begin{aligned}
i_1(t) &= \left| aX(t)+bY(t) \right|^2 \\
i_2(t) &= \left| \left( aX(t)+bY(t) \right) \otimes h(t) \right|^2 \\
i_3(t) &= \left| bX(t)+bY(t) \right|^2 \\
i_4(t) &= \left| \left( bX(t)+bY(t) \right) \otimes h(t) \right|^2 \\
i_5(t) &= \left| bX(t)+aY(t) \right|^2 \\
i_6(t) &= \left| \left( bX(t)+aY(t) \right) \otimes h(t) \right|^2,
\end{aligned} \tag{S2}$$

where $h(t)$ is the time-domain impulse response of dispersive element, and '$\otimes$' is the convolution operation. Next, we can reconstruct eight photocurrents ($R_j(t)$, $j$=1~8) using $i_1(t)$ to $i_6(t)$ and a polarization-independent transformation matrix.

$$\begin{bmatrix} 0 & 0 & 3 & 0 & 0 & 0 \\ 0 & 0 & 0 & 3 & 0 & 0 \\ 2 & 0 & -1 & 0 & 2 & 0 \\ 0 & 2 & 0 & -1 & 0 & 2 \\ 1-\sqrt{3} & 0 & 1 & 0 & 1+\sqrt{3} & 0 \\ 0 & 1-\sqrt{3} & 0 & 1 & 0 & 1+\sqrt{3} \\ 1+\sqrt{3} & 0 & 1 & 0 & 1-\sqrt{3} & 0 \\ 0 & 1+\sqrt{3} & 0 & 1 & 0 & 1-\sqrt{3} \end{bmatrix} \begin{bmatrix} i_1(t) \\ i_2(t) \\ i_3(t) \\ i_4(t) \\ i_5(t) \\ i_6(t) \end{bmatrix} = \begin{bmatrix} |X(t)+Y(t)|^2 \\ |(X(t)+Y(t))\otimes h(t)|^2 \\ |X(t)-Y(t)|^2 \\ |(X(t)-Y(t))\otimes h(t)|^2 \\ |X(t)+jY(t)|^2 \\ |(X(t)+jY(t))\otimes h(t)|^2 \\ |X(t)-jY(t)|^2 \\ |(X(t)-jY(t))\otimes h(t)|^2 \end{bmatrix} \triangleq \begin{bmatrix} R_1(t) \\ R_2(t) \\ R_3(t) \\ R_4(t) \\ R_5(t) \\ R_6(t) \\ R_7(t) \\ R_8(t) \end{bmatrix} \quad \text{(S3)}$$

$R_j(t)$ ($j$=1~8) can be regarded as the photocurrents of functional units from the four optical fields ($X(t)$+$Y(t)$, $X(t)$-$Y(t)$, $X(t)$+$jY(t)$, $X(t)$-$jY(t)$). These four single-polarization optical fields can be expressed as

$$\begin{aligned}
X(t)+Y(t) &= 2\cos\alpha\cos\theta C + (\cos\alpha e^{j\theta} + \sin\alpha)S_x(t) + (\cos\alpha e^{-j\theta} - \sin\alpha)S_y(t) \\
X(t)-Y(t) &= (-2\sin\alpha + j2\cos\alpha\sin\theta)C + (\cos\alpha e^{j\theta} - \sin\alpha)S_x(t) \\
&\quad - (\cos\alpha e^{-j\theta} + \sin\alpha)S_y(t) \\
X(t)+jY(t) &= [\cos\alpha\cos\theta - \sin\alpha + \cos\alpha\sin\theta \\
&\quad + j(\cos\alpha\cos\theta + \sin\alpha + \cos\alpha\sin\theta)]C \\
&\quad + (\cos\alpha e^{j\theta} + j\sin\alpha)S_x(t) + (j\cos\alpha e^{-j\theta} - \sin\alpha)S_y(t) \\
X(t)-jY(t) &= [\cos\alpha\cos\theta - \sin\alpha - \cos\alpha\sin\theta \\
&\quad + j(\cos\alpha\sin\theta - \sin\alpha - \cos\alpha\cos\theta)]C \\
&\quad + (\cos\alpha e^{j\theta} - j\sin\alpha)S_x(t) - (j\cos\alpha e^{-j\theta} + \sin\alpha)S_y
\end{aligned} \quad \text{(S4)}$$

The carrier-to-signal power ratios (CSPR) of them are the functions of two variables $\alpha$ and $\theta$, calculated as

$$CSPR_{X(t)+Y(t)}(\alpha,\theta) = \frac{P\{2\cos\alpha\cos\theta C\}}{P\{(\cos\alpha e^{j\theta}+\sin\alpha)S_x(t)+(\cos\alpha e^{-j\theta}-\sin\alpha)S_y(t)\}}$$

$$= 2\cos^2\alpha\cos^2\theta\cdot C_{th}$$

$$CSPR_{X(t)-Y(t)}(\alpha,\theta) = \frac{P\{(-2\sin\alpha+j2\cos\alpha\sin\theta)C\}}{P\{(\cos\alpha e^{j\theta}-\sin\alpha)S_x(t)-(\cos\alpha e^{-j\theta}+\sin\alpha)S_y(t)\}}$$

$$= (2-2\cos^2\alpha\cos^2\theta)\cdot C_{th}$$

$$CSPR_{X(t)+jY(t)}(\alpha,\theta) = \frac{P\{[\cos\alpha\cos\theta-\sin\alpha+\cos\alpha\sin\theta+j(\cos\alpha\cos\theta+\sin\alpha+\cos\alpha\sin\theta)]C\}}{P\{(\cos\alpha e^{j\theta}+j\sin\alpha)S_x(t)+(j\cos\alpha e^{-j\theta}-\sin\alpha)S_y(t)\}}$$

$$= (1+2\cos^2\alpha\cos\theta\sin\theta)\cdot C_{th}$$

$$CSPR_{X(t)-jY(t)}(\alpha,\theta) = \frac{P\{[\cos\alpha\cos\theta-\sin\alpha+\cos\alpha\sin\theta+j(\cos\alpha\cos\theta+\sin\alpha+\cos\alpha\sin\theta)]C\}}{P\{(\cos\alpha e^{j\theta}-j\sin\alpha)S_x(t)-(j\cos\alpha e^{-j\theta}+\sin\alpha)S_y(t)\}}$$

$$= (1-2\cos^2\alpha\cos\theta\sin\theta)\cdot C_{th}$$

$$\text{(S5)}$$

$P$ is the mean power calculated by mathematical expectation function $E\{\cdot\}$ as

$$P\{S_x(t)\} = E\left\{|S_x(t)|^2\right\} \tag{S6}$$

It can be found that there are always two CSPR values satisfying the perturbation assumption (CSPR > $C_{th}$) for optical field recovery regardless of polarization rotation. For example, when $\alpha$ is $\pi/4$ and $\theta$ is $3\pi/4$, $CSPR_{X(t)+Y(t)}$, $CSPR_{X(t)-Y(t)}$, $CSPR_{X(t)+jY(t)}$, and $CSPR_{X(t)-jY(t)}$ are $0.5\cdot C_{th}$, $1.5\cdot C_{th}$, $0.5\cdot C_{th}$, and $1.5\cdot C_{th}$, respectively. Thus, the single polarization optical field $X(t)$-$Y(t)$ and $X(t)$-$jY(t)$ can be digitally recovered using the optical field linearization and SSBI cancellation algorithm [2-5]. Next, the transmitted Jones vector can be acquired using

$$\begin{bmatrix} C+S_x(t) \\ C+S_y(t) \end{bmatrix} = \begin{bmatrix} \cos\alpha e^{j\theta} & -\sin\alpha \\ \sin\alpha & \cos\alpha e^{-j\theta} \end{bmatrix}^{-1} \begin{bmatrix} 1 & -1 \\ 1 & -j \end{bmatrix}^{-1} \begin{bmatrix} X(t)-Y(t) \\ X(t)-jY(t) \end{bmatrix} \tag{S7}$$

Thus, the dual-polarization signals $S_x(t)$ and $S_y(t)$ are recovered. In this work, a deep neural network is introduced not only as a linear adaptive MIMO equalizer, but also as a nonlinear signal-signal beat interference (SSBI) compensator.

## Supplementary Note 2: Proof for Jones space optical field recovery scheme with arbitrary state of polarization of carrier.

To illustrate that the arbitrary state of polarization (SOP) of the carrier at the transmitter side is compatible with the Jones space optical field recovery (JSFR) scheme, we use $\xi$ *and* $\Psi$ to represent the rotation angle and differential phase between the carriers and signals, respectively. The transmitted Jones vector is given as

$$\left|E_{Tx}\right\rangle = \begin{bmatrix} C + \cos\xi e^{j\Psi} S_x(t) - \sin\xi S_y(t) \\ C + \sin\xi S_x(t) + \cos\xi e^{-j\Psi} S_y(t) \end{bmatrix} \triangleq \begin{bmatrix} C + G_x(t) \\ C + G_y(t) \end{bmatrix} \tag{S8}$$

The compound signals $(\cos\xi e^{j\Psi} S_x(t) - \sin\xi S_y(t), \; \sin\xi S_x(t) + \cos\xi e^{-j\Psi} S_y(t))$ are defined as generalized signals $G_x(t)$ and $G_y(t)$. It is easy to prove that the powers of $G_x(t)$ and $S_x(t)$ are equal, shown as

$$\begin{aligned} P(G_x) &= P\left\{\cos\xi e^{j\Psi} S_x(t) - \sin\xi S_y(t)\right\} = \cos^2\xi P\left\{S_x(t)\right\} + \sin^2\xi P\left\{S_y(t)\right\} = P\left\{S_x(t)\right\} \\ P(G_y) &= P\left\{\sin\xi S_x(t) + \cos\xi e^{-j\Psi} S_y(t)\right\} = \sin^2\xi P\left\{S_x(t)\right\} + \cos^2\xi P\left\{S_y(t)\right\} = P\left\{S_x(t)\right\} \end{aligned} \tag{S9}$$

Herein, the cross terms are cancelled because $S_x(t)$ and $S_y(t)$ are independent identically distributed (i.i.d.) variables with zero mean. Thanks to the properties of unitary matrix, for this new Jones vector $\left|E_{Tx}\right\rangle = \begin{bmatrix} C + G_x(t), C + G_y(t) \end{bmatrix}$, the signal parts $G_x(t)$ and $G_y(t)$ are uncorrelated and orthogonal because the cross-correlation function of them is zero, expressed as

$$\begin{aligned} R_{G_x,G_y}(\tau) &= E\left\{G_x(t)G_y^*(t+\tau)\right\} \\ &= E\left\{[\cos\xi e^{j\Psi} S_x(t) - \sin\xi S_y(t)][\sin\xi S_x^*(t+\tau) + \cos\xi e^{j\Psi} S_y^*(t+\tau)]\right\} \\ &= E\left\{[\cos\xi \sin\xi e^{j\Psi} S_x(t) S_x^*(t+\tau)\right\} - E\left\{[\sin\xi \cos\xi e^{j\Psi} S_y(t) S_y^*(t+\tau)\right\} \\ &= 0 \end{aligned} \tag{S10}$$

So, the cross-correlation coefficient of them equals to zero.

$$\rho_{Gx,Gy} = \frac{R_{Gx,Gy}(\tau) - m_{Gx} m_{Gy}}{\sigma_{Gx} \cdot \sigma_{Gy}} = 0 \tag{S11}$$

The rest of the derivation process is the same as Supplementary Note 1. Once the optical field is recovered in Jones space, $S_x(t)$ and $S_y(t)$ can be obtained from $G_x(t)$ and $G_y(t)$ with adaptive equalization. Therefore, the SOP of the optical carrier requires no alignment at the transmitter for JSFR, as the settings in this proof-of-concept experiment.

## Supplementary Note 3: Detailed experimental setup and device parameters.

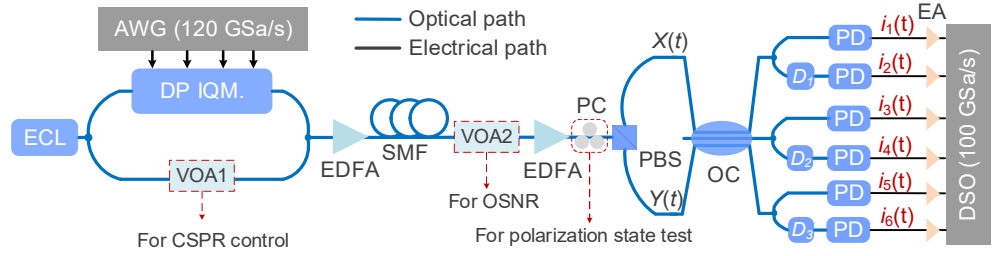

**Figure S1.** Experimental setup. ECL: external cavity laser; AWG: arbitrary waveform generator; DP IQM.: dual-polarization IQ modulator; VOA: variable optical attenuator; EDFA: erbium-doped fiber amplifier; SMF: single-mode fiber; PC: polarization controller; PBS: polarization beam splitter; OC: 3×3 optical coupler; PD: photodiode; EA: electrical amplifier; DSO: digital storage oscilloscope. $D_1$, $D_2$ and $D_3$ are the three dispersive elements with a dispersion value of -300 ps/nm. $i_k(t)$ ($k$=1~6), are the six received photocurrents.

Figure S1 illustrates the detailed setup for four-dimensional modulation and detection experiment. At the transmitter, an arbitrary waveform generator (AWG, Keysight M8194A) with a 3-dB bandwidth of 45 GHz is used to generate four-channel I/Q electrical signals, operating at a sampling rate of 120 GSa/s. The laser source is an external cavity laser (ECL) with a linewidth of ~100 kHz, operating at a wavelength of 1551.06 nm and a launch power of 13 dBm. The integrated dual-polarization IQ modulator (Neophotonics) with a 3-dB bandwidth of 35 GHz modulates the light from the upper branch. Another branch from ECL is used as the optical carrier whose SOP is uncontrolled, satisfying the theoretical analyses in Supplementary Note 2. Herein, the variable optical attenuator (VOA1) is used to adjust the CSPR to satisfy the condition of optical field recovery. After the optical signal and carrier are combined using a 30:70 optical coupler, the dual-polarization optical field is first amplified using an erbium-doped fiber amplifier (EDFA) with a noise figure of about 5 dB and then launched into the fiber link. The launch power is optimized as 8 dBm.

We use G.652 single-mode fiber as the transmission link in the experiment. The typical parameters of attenuation, chromatic dispersion, and polarization mode dispersion are about 0.2 dB/km, 17 ps/(nm·km), and 0.04 ps/$\sqrt{\text{km}}$, respectively.

After the fiber link, the signal is pre-amplified to 18 dBm using another EDFA. Here, the optical signal-to-noise ratio (OSNR) and SOPs of optical signal are adjusted for testing the system performance by VOA2 and polarization controller (PC), respectively, which are not required for the 4-D JSFR scheme. Then, the proposed 4-D JSFR receiver is used to detect the dual-polarization optical field. Here, the 3×3 optical coupler and three two-branch functional units are used to eliminate the polarization fading effect and recover the single-polarization optical field, respectively. Three dispersive elements $D_i$, $i$=1~3, used here are two Fiber Bragg grating (FBG)-based dispersion compensation modules (TDCMX-SM TERAXION) and one wave-shaper (II-V waveshaper 4000A) due to the lack of 3 FBGs. The applied dispersion values are set as about -300 ps/nm. After the optical fields are detected by six single-ended photodetectors (XPD3120R), a real-time digital storage oscilloscope (DSO, Tektronix DPO75902SX) with 6 synchronized channels and 5 effective number of bits, operating at 100 GSa/s, is used to capture the electrical waveforms for off-line DSP. The electrical amplifiers with a 23-dB gain (SHF S807C) are used to suppress the electrical and quantization noise from the DSO.

Note that the bandwidth of JSFR receiver is strictly set as 18.15 GHz via a digital rectangular filter in the receiver-side digital signal processing.

**Supplementary Note 4: Simplified frame structure and fast convergence via transfer learning.**

To reduce the number of training symbols and epochs, transfer learning (TL) has been introduced and proved for speeding up the OSNR estimation, fast remodeling, and nonlinear equalization in optical fiber communications [5].

For the polarization tracking and nonlinearity SSBI elimination of four-dimensional direct detection system in this work, TL can speed up the learning procedure and simplify the frame structure by adjusting the weights of neurons based on the prior knowledge instead of random initialization. TL works efficiently since most of the physical parameters between the source and target domain are almost the same such as the chromatic dispersion values of fiber link and dispersive element, and the frequency response of transceiver. The slow-varying polarization state, in the form of SSBI components to be eliminated, can be tracked using a few training symbols.

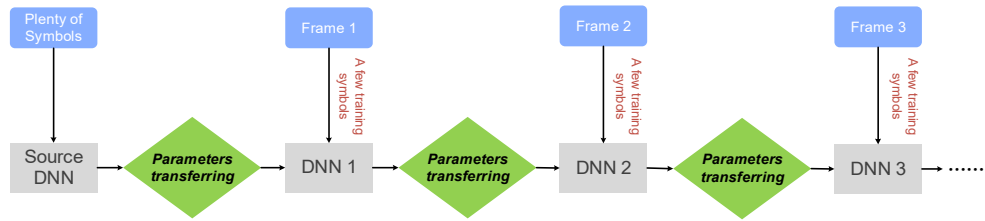

**Figure S2.** The schematic diagram of transfer learning-assisted DNN method for fast remodeling. DNN: deep neural network.

Figure S2 shows the schematic of transfer learning-assisted deep neural network (DNN) method. First, the source DNN is trained using plenty of training symbols (60000 in this experiment) to ensure the convergence. Second, the parameters of source DNN including the weights and bias values are transferred to DNN 1 and thus DNN 1 is trained to realize fast channel remodeling using a few training symbols. Third, the DNN 1 is used to perform optical field linearization for the payload data of Frame 1. Last, the parameters of the current DNN are transferred to the next DNN frame by frame.

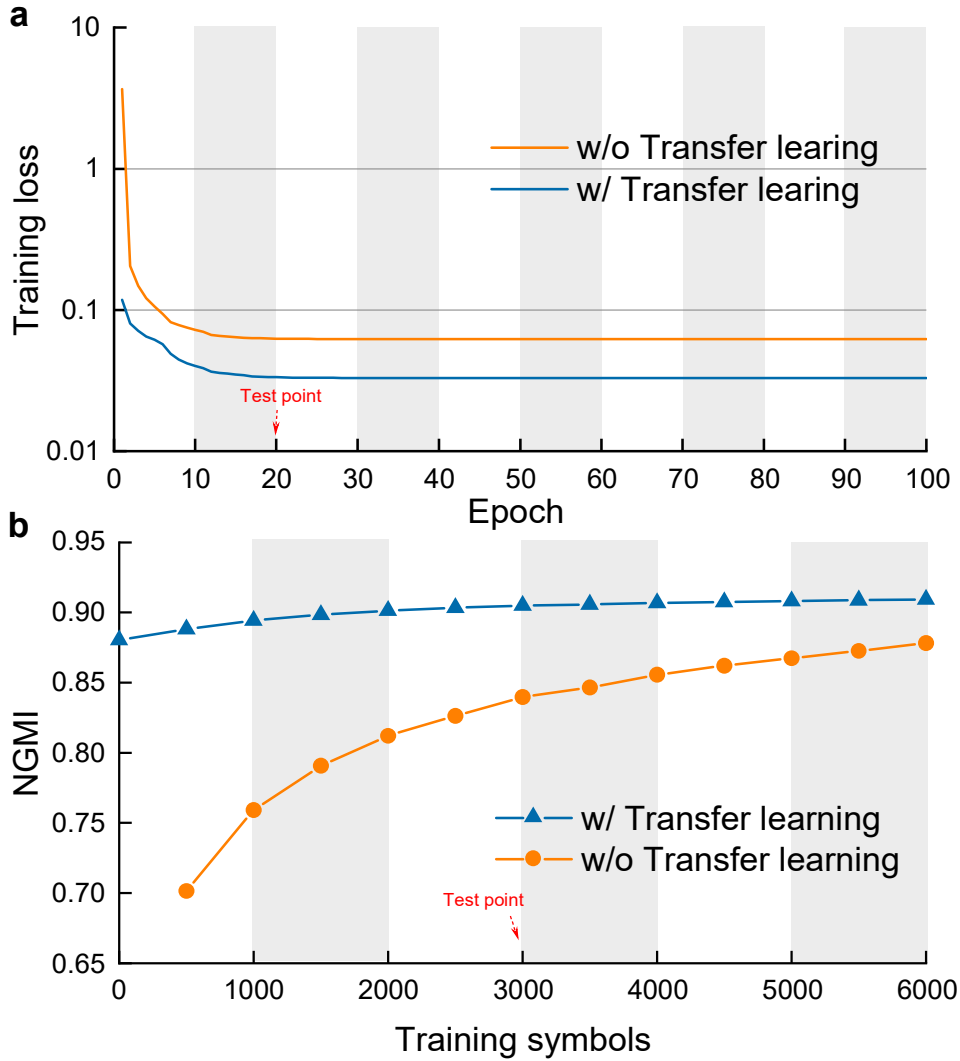

**Figure S3.** (a) The training process using a training dataset of 3000 symbols with and without transfer learning. (b) Measured NGMI as a function of training symbols with and without transfer learning.

In this experiment, we use a few training symbols to realize fast channel remodeling and polarization tracking. The training loss with and without transfer learning versus retraining epoch is shown in Fig. S3(a). The training dataset has 3000 symbols. Thanks to the prior knowledge from the previous frame, TL-assisted DNN can achieve a lower training loss, compared with the conventional DNN without parameters transferring. After the training with 20 epochs, the optimized DNN is switched to test the system performance.

Then, we investigate the normalized mutual information (NGMI) as a function of the training symbols with a TL-assisted DNN, as shown in Fig. S3(b). TL-assisted DNN recover the optical field approximately even without training symbols. This implies that the optical fiber channels between the adjacent frames are highly correlated. To reduce the redundancy of frame, we adopt the TL-assisted DNN with 3000 training symbols for NGMI prediction.

**Supplementary Note 5: Data rate and electrical spectral efficiency calculations with probabilistic shaping.**

We use a probabilistically-shaped 64-ary quadrature amplitude modulation (PS-64QAM) with varying shaping parameter $\beta$ to adjust the entropy of transmitted symbols. The generalized mutual information (GMI) under bit-metric decoding offers a reliable estimation of the lower bound of achievable information rate, which can be estimated using the bit-wise log-likelihood ratio (see Eqs. (8-11) in Ref. [7]).

NGMI is defined as the maximum number of information bits per transmitted bit, can be calculated as

$$NGMI = 1 - [H - GMI] / \log_2(M) \tag{S12}$$

where $H$ is the source entropy, and $M$ the is constellation size.

In this experiment, the entropy of transmitted PS-64QAM symbol is optimized as 4.4 bits/symbol to maximize the GMI. The shaping parameter $\beta$ is 1.1976. A concatenated forward error correction (FEC) composed of an inner spatially-coupled low-density parity-check code and an outer hard-decision BCH code [8] is used to remove the potential error floor. The concatenated FEC has a code rate $R_c$ of 0.8402 and NGMI threshold of 0.8798. $\gamma$ can be calculated as

$$\gamma = 1 - \frac{\log_2(M)}{2} \times (1 - R_c) = 1 - 3 \times (1 - 0.8402) \tag{S13}$$

Therefore, the line rate $R_{Line}$ is 263.8 Gb/s, which is calculated as

$$R_{Line} = 2 \times (1 + \beta) \times r_c \times 2\,pol = 2 \times (1 + 1.1976) \times 30 \times 2 = 263.8 \tag{S14}$$

where $r_c$ is 30(=15×2) GBd for two digital subcarriers modulation on each polarization. The net data rate $R_{info}$ is 206.2 Gb/s, which is calculated as

$$R_{info} = 2 \times (\gamma + \beta) \times r_c \times 2\,pol = 2 \times (0.5206 + 1.1976) \times 30 \times 2 = 206.2 \tag{S15}$$

The emulated receiver bandwidth using a digital rectangular filter is 18.15 GHz. Therefore, the achieved net electrical spectral efficiency ($ESE_{net}$) is 11.36 b/s/Hz, which is calculated as

$$ESE_{net} = R_{info} / (r_c \times (1 + \text{roll-off}) + \text{Guard band}) = 11.36 \tag{S16}$$

where the roll-off factor of the raised cosine filter is 0.01. The guard band is fixed as 3 GHz to mitigate the singularity of transfer function [4]. To the best of our knowledge, it achieves the highest electrical spectral efficiency among the four-dimensional direct detection systems.

## Supplementary References